\documentclass[epj]{svjour}
\usepackage{graphics,amsmath,amssymb}
\begin{document}

\title{Electroweak Physics at LHC}

\author{Jens Erler\thanks{Presented at the 3rd International Workshop "From 
Parity Violation to Hadronic Structure and more..." (PAVI06), Milos Island, 
Greece, May 16--20, 2006.}
}

\institute{Instituto de F\'\i sica, Universidad Nacional Aut\'onoma de 
M\'exico, 04510 M\'exico D.F., M\'exico}

\date{}

\abstract{The prospects for electroweak physics at the LHC are reviewed focusing mainly on precision studies. This includes projections for measurements of the effective $Z$ pole weak mixing angle, $\sin^2\theta_W^{\rm eff.}$, of top quark, $W$ boson, and Higgs scalar properties, and new physics searches.}


\maketitle

\section{Introduction}
\label{intro}

The Large Hadron Collider (LHC) is well on its way to produce first collisions in 2007.  Initial physics runs are scheduled for 2008 with several ${\rm fb}^{-1}$ of data and the precision program can be expected to take off in 2009. The low luminosity phase with about 10~fb$^{-1}$ of data (corresponding to 150 million $W$ bosons, 15 million $Z$ bosons, and 11 million top quarks) per year and experiment~\cite{Pralavorio:2005ik} will already allow most precision studies to be performed. Some specific measurements, most notably competitive results on $\sin^2\theta_W^{\rm eff.}$, will probably have to wait for the high luminosity phase with ${\cal O}(100~{\rm fb}^{-1})$ per year and experiment. The determination of the Higgs self-coupling would even call for a luminosity upgrade by another order of magnitude. 

Good knowledge of the lepton and jet energy scales will be crucial. Initially these will be known to 1\% and 10\%, respectively, but with sufficient data one can use the $Z$ boson mass for calibration, allowing 0.02\% and 1\% determinations. Furthermore, a 2\% measurement of the luminosity and 60\% $b$-tagging efficiency can be assumed~\cite{Pralavorio:2005ik}. 

To give a point of reference, a combination of all currently available precision data yields for the Higgs mass, 
$$ M_H = 88^{+34}_{-26} \mbox{ GeV,}$$
and for the strong coupling, $\alpha_s (M_Z) = 0.1216 \pm 0.0017$. The fit value for the top quark mass, $m_t = 172.5 \pm 2.3$~GeV, is dominated by and coincides with the Tevatron combination~\cite{TevatronEWG:2006}. The $\chi^2/{\rm d.o.f.}$ at the minimum of the global fit is 47.4/42 with a probability for a larger $\chi^2$ of 26\%. The 90\% CL range for $M_H$ is 
$47~\mbox{GeV} < M_H < 146$~GeV, where upon inclusion of direct search results from LEP~2~\cite{Barate:2003sz} the 95\% CL upper limit increases to 185~GeV. Besides the notorious list of 1.5 to 3~$\sigma$ deviations, the electroweak Standard Model (SM) remains in very good shape. One of the largest discrepancies is the NuTeV 
result~\cite{Zeller:2001hh} on deep inelastic neutrino scattering ($\nu$-DIS) which comes with several challenging theory issues~\cite{Londergran:PAVI06}.

When discussing future improvements for the key observables, $m_t$, 
$\sin^2\theta_W^{\rm eff.}$, and the $W$ boson mass, $M_W$, it is useful to keep some benchmark values in mind. An increase of $M_H$ from 100 to 150~GeV (distinguishing between these values provides a rough discriminator between minimal supersymmetry and the SM) is equivalent to a change in $M_W$ by $\Delta M_W = - 25$~MeV. On the other hand, this 25~MeV decrease can be mimicked by $\Delta m_t = - 4$~GeV, and also by an increase of the fine structure constant at the $Z$ scale, $\Delta\alpha(M_Z) = + 0.0014$. We know $\alpha(M_Z)$ an order of magnitude better than this --- despite hadronic uncertainties in its relation to the fine structure constant in the Thomson limit. On the other hand, improving $m_t$ will be important. The same shift in $M_H$ is also equivalent to $\Delta\sin^2\theta_W^{\rm eff.} = + 0.00021$, which in turn can be mimicked by $\Delta m_t = - 6.6$~GeV or by 
$\Delta\alpha(M_Z) = + 0.0006$. Once the Higgs boson has been discovered and its mass determined kinematically, these observables are then free to constrain heavy new particles which cannot be produced or detected directly. As an example serves the mass of the heavier top squark eigenstate in the minimal supersymmetric standard model at certain parameter values~\cite{Erler:2000jg}.

\begin{figure*}[t]
\resizebox{0.99\textwidth}{!}{
\includegraphics{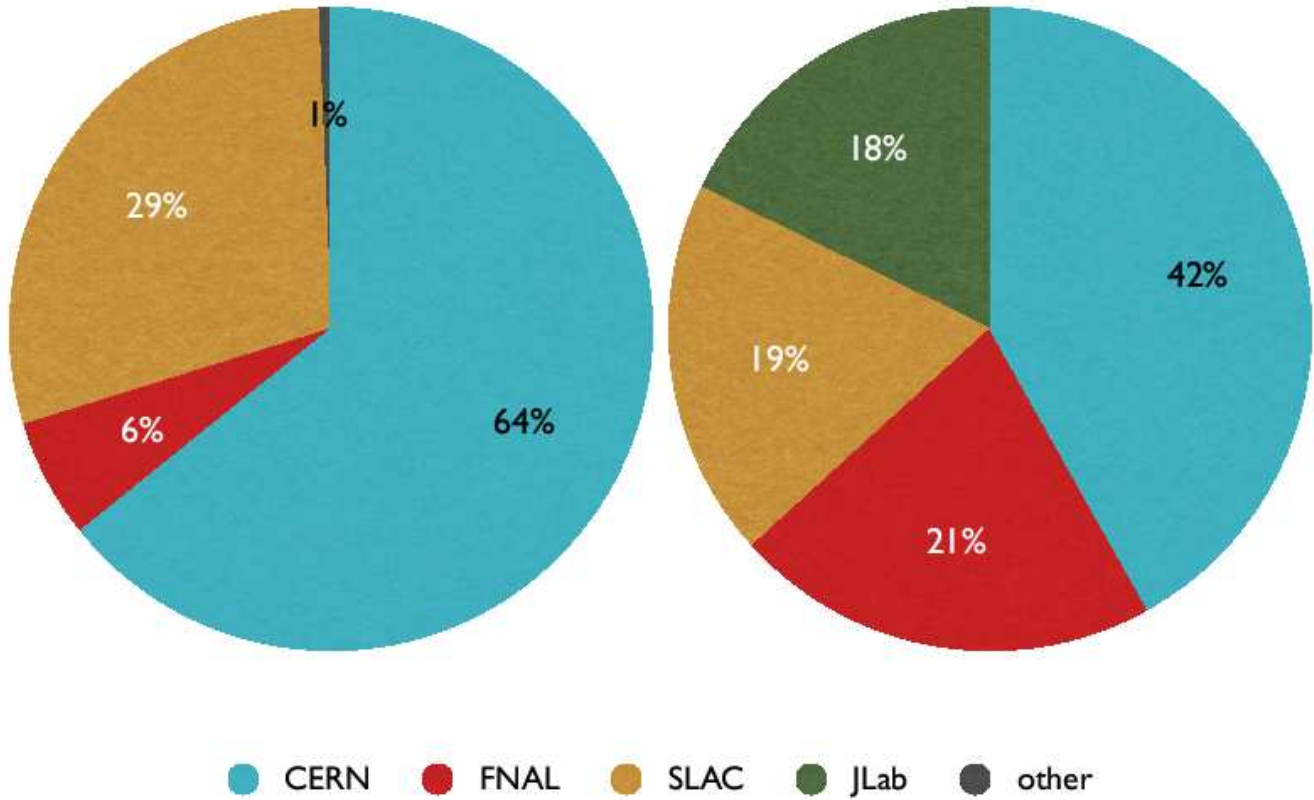}}
\caption{Precision (square of inverse error) weighted contributors to our knowledge of $\sin^2\theta_W^{\rm eff.}$ by laboratory. Shown is the status (left) and a projection (right) for the Tevatron Run IIA (2~${\rm fb}^{-1}$ per experiment) along with 4\% $ep$ and 3.25\% $ee$ cross section asymmetries from JLab.}
\label{s2wbyLab}
\end{figure*}

\section{High precision measurements}

LEP and SLC~\cite{zpole:2005} almost completely dominate the current average $Z$ pole weak mixing angle, $\sin^2\theta_W^{\rm eff.} = 0.23152 \pm 0.00016$. Via measurements of leptonic
forward-backward (FB) asymmetries, the Tevatron Run II is expected to add another combined $\pm 0.0003$ 
determination~\cite{Baur:2005rx}, competitive with the most precise measurements from LEP (the FB asymmetry for $b\bar{b}$ final states) and SLD (the initial state polarization asymmetry). Having 
$p\bar{p}$ collisions are a crucial advantage here. At the LHC, by contrast, one has to focus on events with a kinematics suggesting that a valence quark was involved in the collision and which proton provided it ($Z$ rapidity tag). This will be possible for a small fraction of events only, requiring high luminosity running. Furthermore, sufficient rapidity coverage of $|\eta| < 2.5$ will be necessary for even a modest $\pm 0.00066$ determination~\cite{Quayle:2004ep}. Incidentally, a similar precision is expected from fixed target elastic proton scattering by the Qweak experiment~\cite{Page:PAVI06} using the polarized electron beam at JLab. A breakthrough measurement at the LHC with an error as small as 
$\pm 0.00014$~\cite{Baur:2005rx,Quayle:2004ep} will require a much more challenging rapidity coverage of $|\eta| < 4.9$ for jets and missing transverse energy. Thus, it is presently unclear what the impact of the LHC on $\sin^2\theta_W^{\rm eff.}$ will be. On the other hand, as has been covered at this meeting, there may be further opportunities at JLab after the 12~GeV upgrade of CEBAF in parity violating deep inelastic scattering (DIS-Parity)~\cite{Souder:PAVI06} building on the current 6~GeV DIS-Parity effort~\cite{Zheng:PAVI06}, and most notably, from an improved measurement of polarized M\o ller scattering (e2ePV)~\cite{Mack:PAVI06}, reducing the error of the E~158 experiment at SLAC~\cite{Kumar:PAVI06} by about a factor of four (see Fig.~\ref{s2wbyLab}).

\begin{figure*}[t]
\resizebox{0.99\textwidth}{!}{
\includegraphics{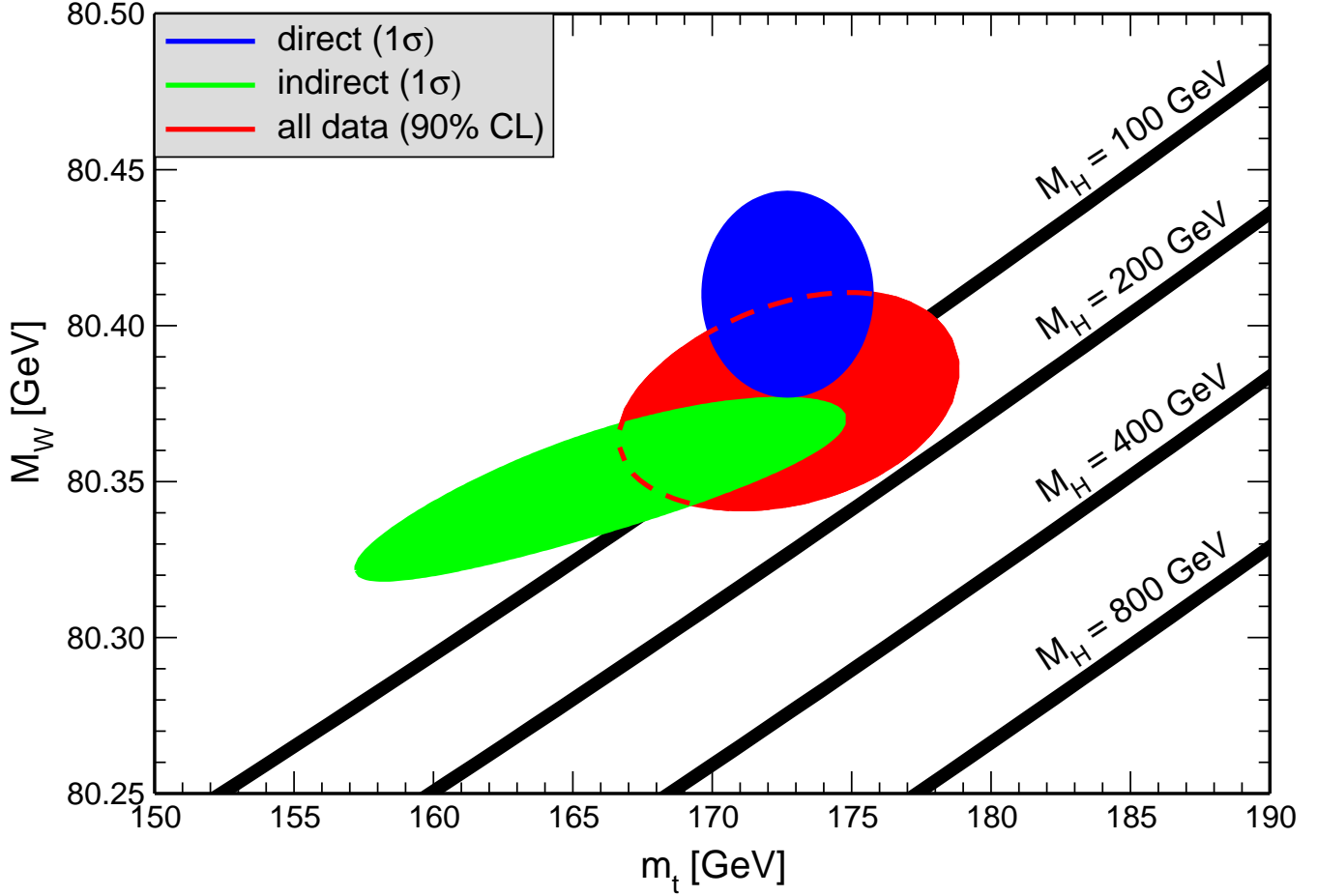}}
\caption{1~$\sigma$ constraints and the 90\% CL allowed region by all precision data in the $M_W-m_t$ plane. The lines show the SM predictions for various Higgs masses, their widths indicating the theory uncertainty from $\alpha(M_Z)$ in the prediction.}
\label{mwmt}
\end{figure*}

Our current knowledge regarding $M_W$ and $m_t$ is summarized in 
Fig.~\ref{mwmt}. The height of the uppermost (blue) ellipse is the average, 
$M_W = 80.410 \pm 0.032$~GeV, of final UA2~\cite{Alitti:1991dk} and Tevatron 
Run~I~\cite{Affolder:2000bp,Abazov:2002bu,Abazov:2003sv}, as well as preliminary LEP~2~\cite{LEP2:2005di} results. With the exception of a less precise threshold determination at LEP~2, all these results are based on direct reconstruction\footnote{Frequently, $\nu$-DIS results are also represented as measurements of $M_W$, since this accounts to a good approximation for the $m_t$ dependence in the SM. This is, however, a coincidence and, in general, $\nu$-DIS is affected differently by new physics than $M_W$, and $\nu$ and $\bar{\nu}$ scattering actually provide two independent observables, although $\bar{\nu}$-DIS is usually less accurate.}. The other (green) 1~$\sigma$ ellipse is from all data excluding $M_W$ and 
the Tevatron $m_t$. Its elongated shape arises because one combination is tightly constrained by $\sin^2\theta_W^{\rm eff.}$ while the orthogonal one is from less precise measurements including low energy observables and the partial $Z$ decay width into $b\bar{b}$ pairs~\cite{zpole:2005} (with a very different $m_t$ dependence than other neutral current observables). Fig.~\ref{mwmt} demonstrates that the direct and indirect contours in the $M_W-m_t$ plane are consistent with each other and independently favor small Higgs masses, $M_H \lesssim 150$~GeV. All channels and experiments combined, the Tevatron Run~II will likely add another $\pm 30$~MeV constraint. The huge number of $W$ bosons will enable the LHC to provide further 
$\pm 30$~MeV measurements per experiment and lepton channel ($e$ and $\mu$) for a combined $\pm 15$~MeV uncertainty (it is assumed here that the additional precision that can be gained by cut optimization is compensated approximately by common systematics). The measurements are limited by the lepton energy and momentum scales, but these can be controlled using leptonic $Z$ decays. With 
the even larger data samples of the high luminosity phase, one may alternatively consider the $W/Z$ transverse mass ratio, opening the avenue to a largely independent measurement with an error as low as 
$\pm 10$~MeV~\cite{Baur:2005rx}, for a combined uncertainty about three times smaller than our benchmark of $\pm 25$~MeV. 

\begin{figure*}[t]
\resizebox{0.99\textwidth}{!}{
\includegraphics{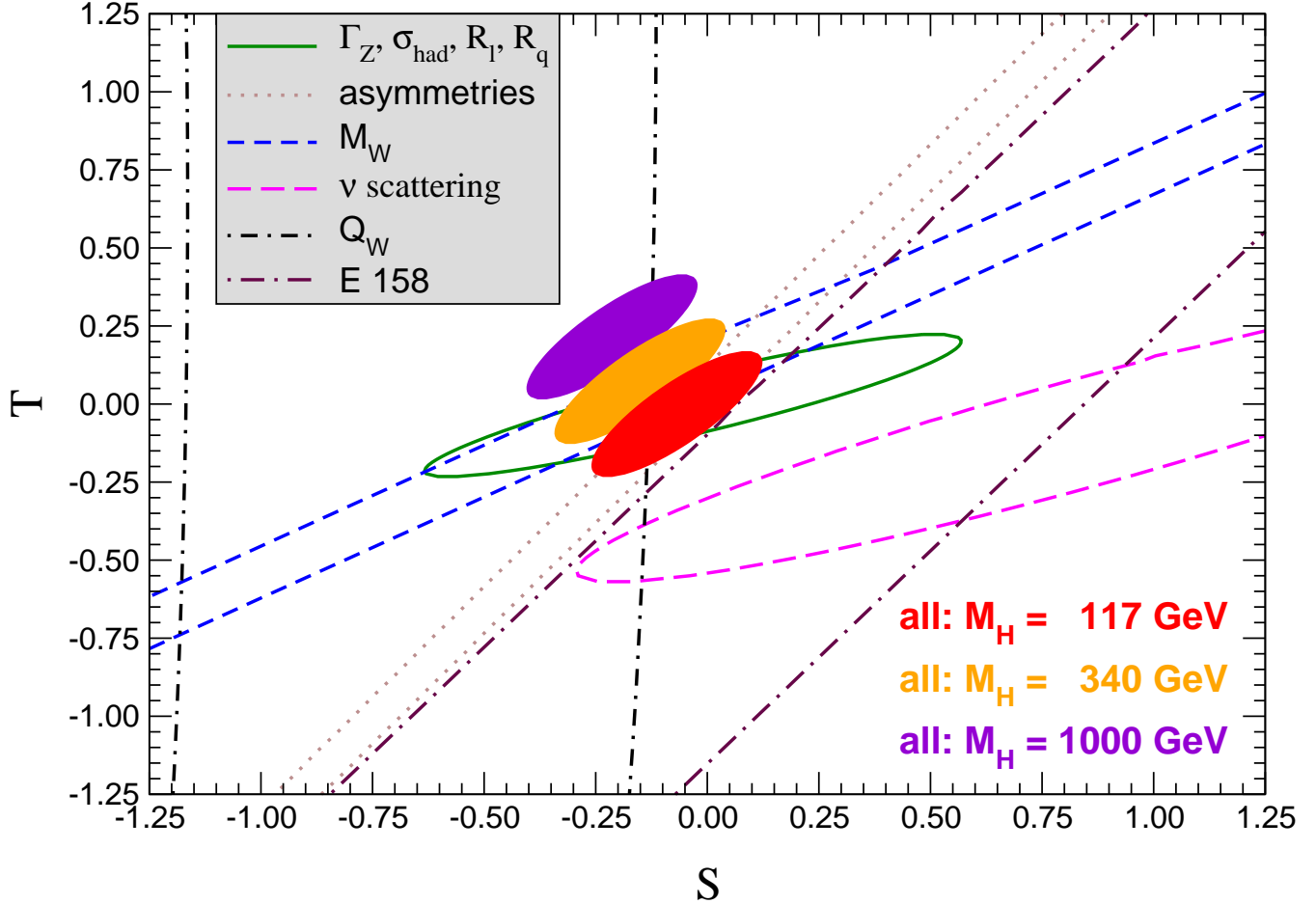}}
\caption{1~$\sigma$ constraints and the 90\% CL regions (for various reference Higgs masses) in $S$ and $T$ allowed by all precision data. They are defined to vanish in the SM, so that any significant deviation from zero may signal the presence of new physics.}
\label{ST}
\end{figure*}

The uncertainty in the world average $\sin^2\theta_W^{\rm eff.}$ after the LHC era will only be half of our benchmark of $\pm 0.00021$ at best, but it is less effected by $m_t$ than $M_W$. More generally, $\sin^2\theta_W^{\rm eff.}$ and $M_W$ can be used to constrain the so-called oblique parameters, describing vector boson self-energies, such as $S$ and $T$~\cite{Peskin:1991sw} shown in Fig.~\ref{ST}. The complementarity between $M_W$ and $\sin^2\theta_W^{\rm eff.}$ (asymmetries and E~158) can be appreciated from the different slopes. Similarly, $Z$ decay properties (other than asymmetries),
$\nu$-DIS, and the weak charges of heavy elements, $Q_W$, from atomic parity violation, all yield different slopes and shapes illustrating the power of having a wide variety of observables available.

The total $W$ decay width, $\Gamma_W$, represents another observable of relevance to oblique parameters, but its sensitivity to new physics and its complementarity to and correlation with other quantities depends on how it is obtained. It can be extracted indirectly through measurements of cross section ratios,
$$
  \left[ 
  {\sigma(pp \to Z \to \ell^+ \ell^- X) \over \sigma(pp \to W \to \ell \nu X)}
  \right]_{\rm exp.} \times
  \left[ 
  {\sigma(pp \to W) \over \sigma(pp \to Z)}
  \right]_{\rm th.} \times
$$
$$
  {\Gamma_{\rm SM}(W\to \ell \nu)\over {\cal B}_{\rm LEP}(Z\to \ell^+ \ell^-)},
$$
(CDF currently quotes 
$\Gamma_W = 2.079 \pm 0.041$~GeV~\cite{Acosta:2004uq}) but the leptonic $W$ decay width has to be input from the SM. More interesting is therefore the direct method using the tail of the transverse mass distribution. An average of final Tevatron Run~I~ and preliminary D\O~II~\cite{Tevatron:2005ij} and 
LEP~2~\cite{LEP2:2005di} results gives, 
$\Gamma_W = 2.103 \pm 0.062$~GeV. The final Tevatron Run~II is expected to contribute $\pm 50$~MeV measurements for each channel and experiment. Detailed studies for the LHC are not yet available, but historically the absolute error in $\Gamma_W$ at hadron colliders has traced roughly the one in $M_W$. If this trend carries over to the LHC, a $\pm 0.5\%$ error of $\Gamma_W$ may be in store. 

Some Tevatron Run~II results are already included in the current 
$\pm 2.3$~GeV~\cite{TevatronEWG:2006} uncertainty in $m_t$, and with the expected total of 8~fb$^{-1}$ the error may decrease by another factor of two. The LHC is anticipated to contribute a 
$\pm 1$~GeV determination from the lepton plus jets channels alone~\cite{Womersley:2006pr}. The cleaner but lower statistics dilepton channels may provide another $\pm 1.7$~GeV determination, 
compared with $\pm 3$~GeV from the systematics limited all hadronic 
channel~\cite{Womersley:2006pr}. The combination of these channels (all dominated by the $b$ jet energy scale) would yield an error close to the additional irreducible theoretical uncertainty of $\pm 0.6$~GeV from the conversion from the pole mass (which is approximately what is being measured~\cite{Smith:1996xz}) to a short distance mass (such as $\overline{\rm MS}$) which actually enters the loops. Folding this in, the grand total may give an error of about $\pm 1$~GeV, so that the parametric uncertainty from $m_t$ in the SM prediction for $M_W$ would be somewhat smaller than the anticipated experimental error in $M_W$.

\section{Other electroweak physics}

With 30~fb$^{-1}$ of data, the LHC will also be able to determine the CKM parameter, $V_{tb}$, in single top quark production to 
$\pm 5\%$~\cite{Mazumdar:2005th} (one expects $\pm 9\%$ from the Tevatron Run~II although no single top events have been observed so far). Anomalous flavor changing neutral current decays, $t\to Vq$ (where $V$ is a gluon, photon, or $Z$ boson, and $q\neq b$), can be searched for down to the $10^{-4}-10^{-5}$ level ~\cite{Womersley:2006pr}. This is a sensitivity gain by three orders of magnitude over current HERA bounds~\cite{Ferrando:2004am}, and relevant, 
{\em e.g.\/} to extra $W'$ bosons. Measuring $t\bar{t}$ spin correlations at the 10\% level~\cite{Womersley:2006pr} will allow to establish the top quark as a spin 1/2 particle, to study non-standard production mechanisms ({\em e.g.\/} through resonances), and to discriminate between $W^+ b$ and charged Higgs ($H^+ b$) decays. 

If the Higgs boson exists, its production at the LHC will proceed primarily through gluon fusion, $gg \to H$, and/or vector boson fusion, $qq' \to H qq'$. Higgs couplings can generally be determined to $10-30\%$~\cite{Baur:2005rx}. The top Yukawa coupling is best studied in associated production, $pp \to t\bar{t} H$, to $20-30\%$ precision~\cite{Womersley:2006pr}. Most difficult proves the Higgs self coupling, $\lambda$, whose measurement would need a luminosity upgrade. With 3~ab$^{-1}$ of data, $\lambda$ can be measured to $\pm 20\%$, for 150~GeV~$< M_H < 200$~GeV, while only 
$\pm 70\%$ precision would be possible for a lighter (and weaker coupled) Higgs boson~\cite{Baur:2005rx}. 

The LHC is, of course, primarily a discovery machine with the outstanding task to find the Higgs boson or else to rule out its existence~\cite{Escalier:2005tq}. As an example for a potential discovery beyond the SM, an extra $Z'$ ($W'$) boson would reveal itself through a high dilepton invariant ($\nu\ell$ transverse) mass peak. Current $Z'$ limits (which depend on the nature of the $Z'$) ranging from 650 to 850~GeV from CDF~\cite{Abulencia:2006iv} and from 434~GeV to 1.8~TeV from LEP~2~\cite{LEP2:2005di} can be extended to 4.2--5~TeV with 100~fb$^{-1}$ of data, while 1~ab$^{-1}$ from an upgraded LHC would add another TeV to the reach~\cite{Godfrey:2002tn}. 

In some cases one can turn things around and use electroweak physics to understand the LHC. For example, by computing $W$ and $Z$ cross sections and comparing them to LHC production rates one can extract the luminosity of the machine. This assumes knowledge of the relevant parton density functions (PDFs) which will probably be available with 2\% uncertainties reflecting the limitation of the method~\cite{Mazumdar:2005th}. In turn, one can obtain information on $u$ and $d$ quark PDFs by measuring the $W^\pm$ charge 
asymmetry, defined as the differential (with respect to the $e^\pm$ rapidity) cross section asymmetry~\cite{Baur:2005rx}.

\section{Conclusions}

The LHC is posed to achieve breakthrough discoveries in the electroweak symmetry breaking sector. As for precision measurements, one can expect particularly great improvements in $M_W$, while a competitive measurement of 
$\sin^2\theta_W^{\rm eff.}$ will be possible with large rapidity coverage only. Measurements of $m_t$, $\Gamma_W$, Yukawa and Higgs self-couplings will be performed and the top quark will be subjected to detailed studies. 

At a meeting on parity violation these measurements should be put into context with low energy precision measurements which will remain important complements even with the LHC in operation. This is because (i) they are capable of contributing results on $\sin^2\theta_W^{\rm eff.}$ which can compete both with $Z$ pole 
factories and hadron colliders (see Fig~\ref{s2wbyLab}), (ii) they are subject to entirely different experimental and theoretical issues, and (iii) they are generally affected quite differently by beyond the SM physics. 

\section*{Acknowledgments}

It is a pleasure to thank the organizers of PAVI06 for the invitation to a very enjoyable meeting. This work was supported by CONACyT (M\'exico) contract 42026--F and by DGAPA--UNAM contract PAPIIT IN112902.

\end{document}